\documentclass[%
 aip,
 jmp,%
 amsmath,amssymb,
 reprint,%
author-numerical,%
onecolumn,%
]{revtex4-1}

\usepackage{graphicx}
\usepackage{dcolumn}
\usepackage{bm}
\usepackage{xcolor}
\usepackage{hyperref}
\usepackage{amssymb} 
\renewcommand{\Bbb}{\mathbb}

\def\al{\alpha}
\def\be{\beta}
\def\de{\delta}
\def\ga{\gamma}

\def\ep{\epsilon}

\def\te{\theta}

\def\ze{\zeta}
\def\om{\omega}
\def\si{\sigma}
\def\vp{\varphi}

\def\Ga{\Gamma}

\def\Om{\Omega}

\def\({\left(}
\def\){\right)}
\def\[{\left[}
\def\]{\right]}

 \def\R{{{\Bbb R}}}

 \def\R{{{\Bbb R}}}

\def\arr{\rightarrow}

\def\then{\Rightarrow}

\def\binomial#1#2{\left(\begin{matrix}#1\cr #2\\\end{matrix}\right)}

\def\Frac[#1/#2]{\frac{#1}{#2}}
\def\del{\partial}

\begin{document}

\preprint{AIP/123-QED}

\title[Discrete rPS]{Discrete Relativistic Positioning Systems}

\author{S. Carloni}
\affiliation{
Centro de Astrofis\'\i ca e Gravita\c{c}\~ao - CENTRA,
Instituto Superior T\'{e}cnico - IST,
Universidade de Lisboa - UL,
Avenida Rovisco Pais 1, 1049-001, Portugal
}%
\email{sante.carloni@gmail.com}

\author{L.Fatibene}
\affiliation{%
Department of Mathematics, University of Torino, Torino 10123, Italy
}%
\affiliation{%
INFN Torino - Iniziativa Specifica QGSKY
}%
\altaffiliation[L.F. on sabbatical leave to ]{Department of Applied Mathematics, University of Waterloo (ON Canada)}
\email{lorenzo.fatibene@unto.it}

\author{M.Ferraris}
\affiliation{%
Department of Mathematics, University of Torino, Torino 10123, Italy}%
\email{marco.ferraris@unito.it}

\author{R.G. McLenaghan}
\affiliation{
Department of Applied Mathematics, University of Waterloo (ON Canada)}%
\email{rgmclenaghan@uwaterloo.ca}

\author{P.Pinto}
\affiliation{Physics Department, Lancaster University, Lancaster LA1 4YB,
UK}%
\email{paolopinto91@gmail.com}

\date{\today}

\begin{abstract}
We discuss the design for a discrete, immediate, simple  relativistic positioning system (rPS) which is potentially able of self-positioning (up to isometries) and operating without calibration or ground control assistance.
The design is discussed in {1+1 spacetimes}, in Minkowski and Schwarzschild solutions, as well as in {2+1 spacetimes} in Minkowski.

The system works without calibration, {i.e.~}clock synchronizations, or {\it prior} knowledge about the motion of clocks, it is {{\it robust}, i.e.~it is able to test} hypotheses break down (for example, if one {or more clocks} temporarily become not-freely falling, or the gravitational field changes), and then it is automatically back and operational when the assumed conditions are restored.

In the Schwarzschild case, we {also check} that the system can best fit the gravitational mass of the source of the gravitational field. {We} stress that no weak field assumptions are made anywhere.
In particular, the rPS we propose can work in a region close to the horizon since it does not use approximations or PPN expansions.
More generally, the rPS can be adapted as detectors for the gravitational field and we shall briefly discuss their role in testing different theoretical settings for gravity.
In fact, rPS is a natural candidate for a canonical method to extract observables out of a gravitational theory, an activity also known as {\it designing experiments to test gravity}.

\end{abstract}

\pacs{Valid PACS appear here}
                              
\maketitle


\section{\label{sec:level1}Introduction}

General Relativity (GR) is, first of all, a framework for defining physical theories in which one can obtain an absolute (i.e.~independent of the observer) description of Nature; {see \cite{Iorio, Smoot, Vishwakarma}}. 
As such it is assumed as the most fundamental framework for describing the classical regime and any description of physical events in that regime should comply {with} it.

Surprisingly, there are a lot of things that can be described in a covariant way (from dynamics to conserved quantities, from some frameworks of Hamiltonian formalism to stability).
Although  this program is carried out in many instances, in many other cases this is not what actually happens. 
For example, astronomers are routinely measuring the positions of stars in the sky, their mutual distances,  the age of the universe, or
the deflection angles nearby massive objects (which are usually thought as ``the angle'' between two spacelike vectors applied at different points in space). 
The NAVSTAR--GPS is paradigmatic of this attitude: it  measures users' positions {\it in space}. 
Moreover, it relies on keeping clocks synchronised despite their motion and the different gravitational potential they experience.
None of these quantities are covariant; see{\cite{Brumberg91,  Brumberg10,  Soffel}}.

In view of this lack in covariance, one can either accept that these quantities depend on conventions (e.g.~protocols to define synchronisation at a distance) and describe in detail
the conventions used, or one can reduce them to measuring coincidences, which are the only absolute quantities one can resort to.
While the first strategy is often tacitly assumed, the second one, which is relativistically more appealing, it is hardly ever tried out in practice (with some exceptions; see 
 \cite{Rovelli}).

Most of the time, we keep assuming to live in a {Newtonian spacetime}, though with some corrections due to what we learned in the last century.
That is particularly evident with NAVSTAR--GPS which was originally designed to work in a Newtonian space with corrections due to GR (including special relativity (SR) corrections, in particular).

Of course, the approximations are very reasonable since the gravitational field of the Earth happens to be weak enough to justify them and this is why NAVSTAR--GPS works well despite its poor theoretical design;  {see \cite{Ash1, PS}}. 
The design of Galileo Global Navigation Satellite System (Galileo--GNSS) better integrates GR, though still on a post-Newtonian regime, 
thus not providing a qualitatively different approach under this viewpoint, e.g.~it still relies on the weak field approximation; {see \cite{Galileo, Galileo2, Galileo3}}.

{Here our approach will be different. We first establish an exact model in a simplified physical situation, with an isotropic gravitational field. This allows an exact treatment with no approximations.  Naturally, in refining the model we shall be forced to add perturbations to the model to describe finer physical effects (non-isotropy, time drift functions of the clocks, dragging forces). However, in this paper we do not approximate or use expansions series, or PN approximations. We just solve the exact equations as we still do not know if these effects will be relevant for the positioning part, or for both the positioning and gravitometric parts. Such evaluation will be left for a later work.}

Recently,  it has been argued {by B.Coll (see \cite{Co1})} that a completely new, qualitatively different,  relativistic design for positioning systems (rPS, {also called {\it emission coordinates}, {\it null coordinates}, {\it light coordinates}, {\it null frames}, or {\it ABC coordinates}}) is needed; see {also \cite{RovelliGPS, Bla,Coll1, Coll2, Coll3, Coll4, Tarantola}}. 
These should be based on a cluster of {\it transmitters}  (or {\it clocks}) broadcasting information with which a {\it user}  (or {\it {client}}) can determine its position in spacetime.
{Unlike in radar coordinates in which a signal goes back and forth between the radar and the object to be located, in positioning systems the object to be located only receives signals from transmitters. Later on we will allow signals to travel among transmitters, though no signals go from the object to the transmitters.}
The main characteristics of a rPS should be:

{\sl
\begin{itemize}
\item[(i)] it should determine the position of events in spacetime, not in space;
\item[(ii)] it should not assume synchronisation at a distance or positioning of initial conditions;
\item[(iii)] it should define a coordinate system which is not linked to Earth leaving it to the {users} the duty to transform it to a more familiar (as well as less fundamental) Earth-based coordinate system.
\end{itemize}
}

Coll {\it et al.~}analysed these rPSs and proposed a classification for them depending on their characteristics.
In their classification, a rPS is {\it generic} if can be built in any spacetime, it is {\it gravity free} if one does not need to know the metric field to built it, it is {\it immediate}
if any event can compute its position in spacetime as soon as it receives the data from transmitters.

Another important characteristic of rPS is being {\it auto-locating}, meaning that the user is able not only to determine its position in spacetime, 
but also the position of the transmitters. This can be achieved by allowing the transmitters to also receive the signals from other clocks and mirroring them together with their clock reading.

The basic design investigated in  \cite{Coll1} is a cluster of atomic clocks based on satellites which {\it continuously } broadcast the time reading of their clock together with the readings they are receiving from the other clocks. 
Sometimes they argue the transmitters may be also equipped with accelerometers or the users can be equipped with a clock themselves.  
In these rPSs the user receives the readings of the transmitters' clock, together with the readings which each transmitter received from the others, for a total of $m^2$ readings for $m$ transmitters.
In these systems, the user may have no {\it a priori} knowledge of transmitter trajectories which are determined by received data 
(sometimes assuming a qualitative knowledge of the kind of gravitational field in which they move or whether they are free falling or subject to other forces).
As a matter of fact, one can define many different settings and investigate what can be computed by the user depending on its {\it a priori} knowledge and assumptions.
For example, it has been shown that these rPS can be used to measure the gravitational field; {see \cite{RovelliGPS, Kostic,Puchades12, Puchades14,  Rovelli, RovelliGPSb, Coll10, Coll10b, Coll12, Coll09, Delva, Bunandar, Bini, Cadez, Gomboc, Lachieze}}.

These rPSs define a family of basic null coordinate systems (in dimension $m$ an event receiving the readings of $m$ clocks can directly use, in some regions of spacetimes, these readings as local coordinates). 
Coll and collaborators showed that one can consider settings so that the rPS is at the same time generic, gravity free and immediate.
The user in these rPSs is potentially able to define familiar (i.e.~more or less related to the Earth) coordinates, as well. 

We should also remark that  there is a rich, sometime implicit, tradition of rPS. It goes back to Ehlers-Pirani-Schild (EPS) who in 1972 proposed an axiomatics for gravitational physics in which the differential structure of spacetime is defined by declaring that {\it radar coordinates} are admissible coordinates; see \cite{EPS, EPSNostro} and \cite{Polistina}.  
Earlier, Bondi and Synge used radar coordinates as somehow preferred coordinates in GR (see \cite{Bondi, WorldFunction, WorldFunction2})
though the tradition goes back to SR as well as before radar was invented (see \cite{Milne1, Milne2, Schroedinger, Whitrow, Whitrow2, Whitrow3}).
Of course, as Coll {\it et al.~}noticed, radar coordinates are not immediate, but, as essentially EPS showed, still they are generic and gravity free.

\medskip

In this paper, we shall {further investigate these issues} proposing an expanded classification of rPS.
In particular, we say that a rPS is {\it chronal} if it only uses clocks, {\it simple} if it is chronal and users have no clock but they uniquely rely on transmitters' clocks.
We also say that a rPS is {\it instantaneous} if a user is regarded as an event, not as a worldline, and it is still able to determine immediately its position in spacetime.
Moreover, we say that a rPS is {\it discrete} if the signals used by the user are a discrete set of clock readings (as opposite to a {\it continuous} stream of them).
We say that a rPS is {\it self-calibrating} if {transmitters starting in generic initial conditions, with clocks which are not synchronised at a  distance, 
are able to operate as a rPS without external assistance. In particular, this means that the user is not assuming any {\it a priori} knowledge about the specific orbits of the transmitters (other than knowing that they are assumed to be freely falling or the kind of forces that may act on them) or about the time at which each clock has been reset.
All these parameters can be added as unknowns and fitted by the user.
Hence being self-calibrating is a combination of being auto-locating and not requiring clock synchronization at a distance, neither it being an initial synchronisation nor  even more so a periodic one. Of course, one can still assume a knowledge about clocks frequencies and their relation with clock proper time which can be obtained before the mission starts.}

{Finally, we call {\it robust} a rPS that is able, by using only signals within the transmitter constellation, to check of all {\it a priori} assumptions (such as, e.g,, the transmitters being freely falling, or the gravitational flied described by a specific, or general, Schwarzschild metric) are valid and pause working as a rPS in order to prevent wrong positioning. Being robust is a prerequisite to allow a software layer able to adapt to transient effects, e.g.~by adding unknown parameters describing perturbations so that they can be determined by a fit.
However, we shall not explore here this possibility leaving it to a further investigation.
Accordingly, an rPS which is self-calibrating and robust can, in principle (of course, depending on the type of perturbation) react to transient effects by pausing and resuming as soon as the operational conditions are restored, without receiving external assistance.
}

We shall discuss some settings which implement simple, instantaneous, discrete and self-calibrating (as well as general, immediate and self-locating) rPS.
We also discuss how the users can explicitly find the coordinate transformation to familiar systems (e.g.~inertial coordinates {$(t, x)$} in Minkowski,
or $(t, r)$ in Schwarzschild) since, even though null coordinates are more fundamental, they are not practically useful for the GPS user
(as well as, doing that, one also proves that those classes of coordinates are also admitted by spacetime differential structure).
{In some of these cases we shall discuss how the design is robust against unexpected perturbations or, more generally, how the system is able to test the assumptions done about transmitters trajectories and clocks.}

Here we argue that being {\it simple} is important from a foundational viewpoint. 
Atomic clocks are already complicated objects from a theoretical perspective. They can be accepted as an extra structure but that does not mean that one should accept
other apparata (e.g.~accelerometers or rulers) as well. {These} sometimes can be defined in terms of clocks, sometimes are even more difficult to be described theoretically,
sometimes, finally, they are simply ill-defined in a relativistic setting (as rulers are).
Moreover, atomic clocks are complex (as well as expensive) technological systems; while it is reasonable to disseminate a small number of them keeping their quality high,
it is not reasonable to impose each {user} to maintain one of them without increasing costs and worsening quality.

Studying {\it instantaneous} and {\it discrete} rPS is interesting because it keeps the information used to define the system finite {at any time}.
Coll and collaborators, for example, describe the clock readings by {continuous} functions {of proper time}. 
This does not really affect the analysis as long as the positioning is done in null coordinates, but it essentially enters into the game when one wants to transform null coordinates into 
more familiar ones (e.g.~inertial coordinates {$(t, x)$} in Minkowski spacetime). 
{Discrete positioning has been considered in the literature, for example in \cite{T1, T2}.}
We have to remark that clocks are essentially and intrinsically discrete objects. 
Regarding them as continuous objects can be done by interpolation, which partially spoils their direct physical  meaning, as well as introduces approximation biases.
As long as possible, also in this case as for simple rPS, we prefer to adhere to simplicity.

Finally, {\it self-calibrating} rPS are a natural extension of self-positioning systems. If one has a self-positioning system there should be no need to trace the 
trajectories of transmitters back in time. 
We believe it is interesting to explicitly keep track of how long back the user needs to know the transmitters{' orbits}, both from a fundamental viewpoint and for later error estimates,
e.g.~in case one wanted or needed to take into account anisotropies of the Earth's gravitational field.
Of course, a self-calibrating rPS is also auto-locating. 
We can assume though that a {robust and} self-calibrating system, if temporarily disturbed by a perturbation (e.g.~a transient force acting for a while), will detect the perturbation and go back 
into operational automatically and with no external action as soon as the perturbation has gone. 
    
\medskip
{We shall now discuss some simple examples with the aim to illustrate methods which can be useful to deal with more realistic situations. In particular, we shall discuss the simple cases of Minkowski space in 1+1  (one spatial plus one time) and 1+3 dimensions.
We will show that limiting to 1+1 dimensions does not play an essential role and we will give an idea on how to scale to higher dimensions.
In Minkowski we already know the form of the general geodesics and we can focus on simple, discrete, instantaneous and self-calibrating rPS.}
We shall also consider the 1+1 Schwarzschild case to check that flatness does not play an essential role.
{In fact, in this case we shall introduce a method based on Hamilton Jacobi complete integrals, which appears to be applicable more generally to higher dimensional spacetimes.}
We point out that any Lorentzian manifold is locally not too different from the corresponding Minkowski space, so that what we do can also be interpreted as a local approximation in the general case.
However, we shall not investigate here for how long such approximations would remain valid.

We also remark that, in Minkowski space (as well as in Schwarzschild) one has Killing vectors, and if one drags the {user} and all the clocks along an isometric flow, then the whole sequence of signals is left invariant. Accordingly, when Killing vectors are present, obviously, one cannot determine the position of anything, since all positions are determined up to an isometry,
and one can use this to set one clock in a given simple form (e.g.~at rest). 

In Section II, we shall consider the simple case of Minkowski in {1+1 dimensions}. That is mainly to introduce notation and better present the main ideas.
In Section III, we consider the extension to 1+2 and 1+3 dimension, with the aim of introducing further complexity, though not curvature, yet.
In Section {IV}, we consider {a} Schwarzschild spacetime in {1+1 dimensions} to check that situations where the curvature is relevant (and consequently, with no affine structure) can be solved as well.
This is done by introducing methods based on symplectic geometry and Hamiltonian framework which appear to be important even in more realistic situations.
Finally, we briefly give some perspective for future investigations.

{We also have two appendices. 
In Appendix A, we briefly discuss how would it be the theory in Minkowski spacetime of arbitrary dimension.
In Appendix B, we briefly discuss the relation between our evolution generator and Synge's world function.
}

\section{Minkowski case in 1+1 dimensions}

{Let us start with a somehow trivial example. We consider positioning in Minkowski in 1+1 dimensions.
Here freely falling worldlines are straight lines. We do not expect Minkowski to be a realistic model of Earth gravitational field, unless we are very far from it.
This example is meant to introduce some ideas to be used later in Schwarzschild which will also be 1+1 dimensional.

}

Let us assume the spacetime $M$ to be flat and 2-dimensional.
Although what we shall discuss is intrinsic, let us use a system of Cartesian coordinates $(t, x)$ to sketch objects.

Since there is no gravitational field, particles {through the event $(t_\ast, x_\ast)$} move along geodesics which are straight lines 
\begin{equation}
x-x_\ast = \be (t-t_\ast)
\qquad
-1< \be < 1
\end{equation}
while for light rays  {through the event $(t_\ast, x_\ast)$} have $|\be|=1$, i.e.
\begin{equation}
x-x_\ast = \pm (t-t_\ast)
\end{equation}

In view of covariance, what we are saying is that free fall is expressed by first order polynomials {\it in the given coordinates $(t, x)$}.
If we use polar coordinates $(r, \te)$, free fall would not be given by first order polynomials (such as $r-r_\ast= \ga(\te-\te_\ast)$).
It would be rather given by the {\it same} straight lines (e.g.~$x-x_\ast = \be (t-t_\ast)$) expressed in the new coordinates, i.e.
\begin{equation}
r \sin(\te)-r_\ast \sin(\te_\ast) = \be \(r \cos(\te)-r_\ast \cos(\te_\ast)\)
\end{equation}
which are in fact the {\it same} curves.

It is precisely because of this fact that here we are not using coordinates in an essential way (and thus not spoiling covariance). 
We instead are just selecting a class of intrinsic curves to represent free fall.

A {\it clock} is a parametrised particle world line 
\begin{equation}
\chi: \R\arr M: s\mapsto (t(s), x(s))
\end{equation} 
A {\it standard clock} is a clock for which the covariant acceleration $a^\mu := \ddot x^\mu+ \Ga^\mu_{\al\be} \dot x^\al \dot x^\be$
 is perpendicular to its covariant velocity $\dot x^\al$ (see \cite{Perlick, Polistina}); in this case, and in Cartesian coordinates, 
the acceleration is given by the second derivative (since $\Ga^\mu_{\al\be}= \{g\}^\mu_{\al\be}$ and Christoffel symbols are vanishing).
Since the clock is moving along a straight line, then it is standard iff the functions $t(s), x(s)$ are linear in $s$.
Hence the most general standard clock {through the event $(t_\ast, x_\ast)$} is 
\begin{equation}
\chi_\ast: \R\arr M: s\mapsto  (t= t_\ast+ \al s, x= x_\ast + \al \be s)
\end{equation} 
Its covariant velocity $\dot \chi_\ast$ is constant and one can always set its rate $\al$ so that $|\dot \chi_\ast|^2=-1$ is normalised.
In that case, one sets $\ze:= \al \be$, so that $|\dot \chi_\ast|^2= -\al^2+ \ze^2=-1$ ($\then\>\al^2- \ze^2=1$), i.e.
\begin{equation}
\al=\Frac[1/\sqrt{1-\ze^2}]
\label{ProperClocks}
\end{equation}
which is called a {\it proper clock}.
A proper clock {measures its proper time $s$ and it} has three degrees of freedom since it is uniquely determined by four parameters $(t_\ast, x_\ast, \al, \ze)$ with the relation (\ref{ProperClocks}).
{Although hereafter we shall consider only proper clocks, it is not difficult to introduce more realistic, {\it a priori} known, drift functions to describe actual atomic clocks.}

Let us consider two proper clocks $(\chi_0, \chi_1)$ in $M$ corresponding to the parameters $(t_0, x_0, \al_0, \ze_0)$ and $(t_1, x_1, \al_1, \ze_1)$.
As we anticipated above the whole system has a Poincar\'e invariance which can be fixed by setting $\chi_0: \R\arr M: s\mapsto  (t= s, x= 0)$
(which still leaves an invariance with respect to spatial reflections, which will be eventually used) and consequently, $\chi_1: \R\arr M: s\mapsto  (t= t_1+ \al s, x= x_1 + \ze s)$.

Before proceeding, let us once again explain which problem we intend to consider in the following.
The usual rPS would assume $(t_1, x_1, \al, \ze)$ to be known parameters fixed during the calibration of the system.
A {\it {user}} receiving the values of $s_i$ (with $i=0,1$) from the clocks at an event  $c=(t_c, x_c)$ is able to compute its position $(t_c, x_c)$ as a function of the signals $(s_0, s_1)$.

The signals $(s_0, s_1)$ are assumed to be coordinates on the spacetime manifolds, and one can prove that the transition functions 
$\vp:(s_0, s_1)\mapsto (t, x)$ are smooth, so that also $(t, x)$ are good coordinates on spacetime.

This case is particularly simple. The situation is described in Figure 1.

\begin{figure}[htbp] 
   \centering
 \includegraphics[width=9cm]{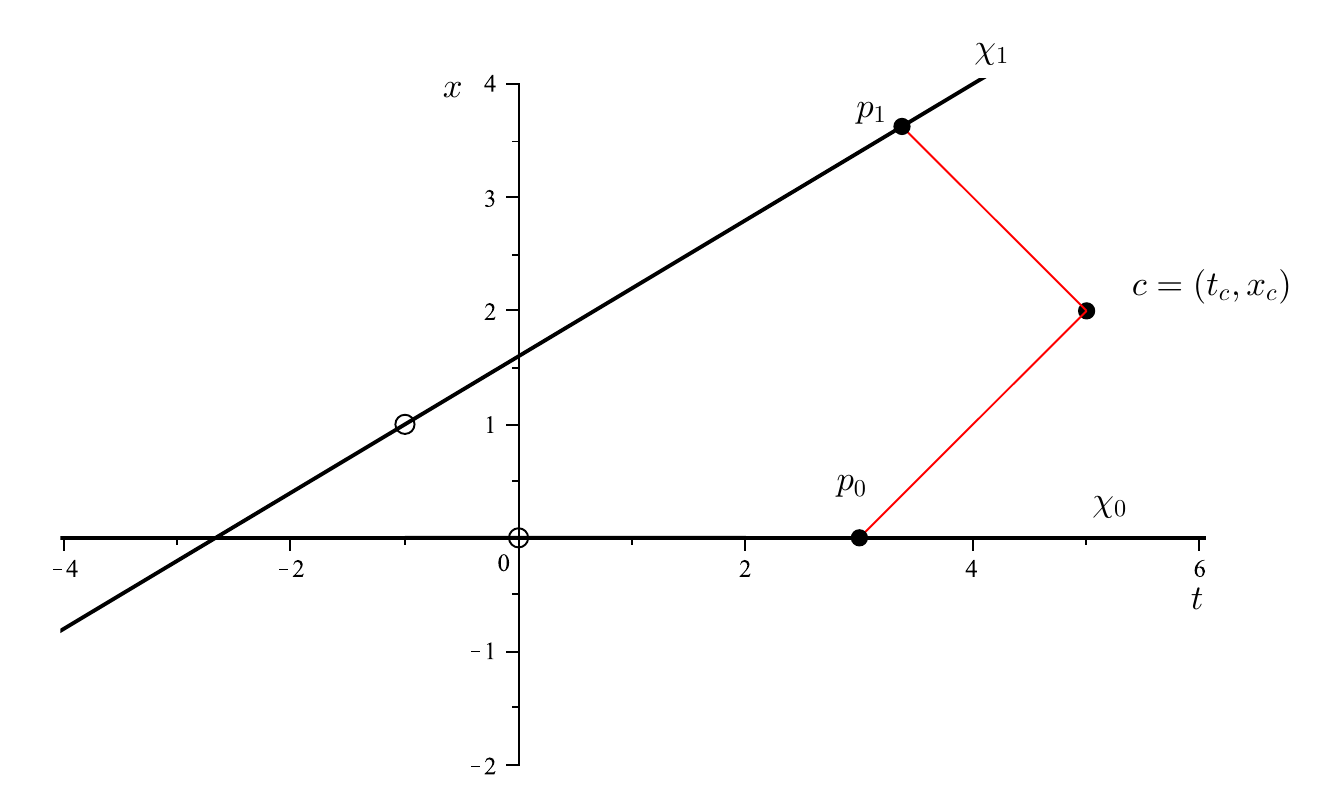} 
   \caption{\small\it  Messages $(s_0, s_1)$ exchanged from the clocks to the {user}. The empty circles are the events when the clocks have been reset.}
   \label{fig:F1}
\end{figure}

One can follow Fig.~1 and readily compute that
\begin{equation}
s_0=t_c-x_c
\qquad
s_1=\Frac[(x_c-x_1) + (t_c-t_1) /\al+\ze]
\end{equation}
which can be readily solved for $(t_c, x_c)$ to get
\begin{equation}
t_c= \Frac[1/2]\[ (\al +\ze )s_1+ s_0+t_1+x_1\]
\qquad\qquad
x_c=\Frac[1/2]\[(\al +\ze )s_1-s_0+t_1+x_1\]
\label{TransFunc}
\end{equation}

Accordingly, one can {\it define} the coordinates $(t, x):=(t_c, x_c)$ as above.
Equations (\ref{TransFunc}) define transition functions between coordinates $(s_0, s_1)$ and $(t, c)$, which are regular being polynomial. We should stress that here $(t_1, x_1, \al,\ze)$ are treated as known parameters.

Our problem in the following, will be to show that if we promote $(t_1, x_1, \al, \ze)$ to be unknowns of the problem together with $(t_c, x_c)$, and we add a whole past sequence of readings
(see Figure 2 below), 
then we are still able to solve the system and use the infinite redundancy to check the assumptions of the model (e.g.~that the gravitational field is vanishing, that the clocks are free falling, that the clocks are identical proper clocks, \dots).

{In Fig.2 below we are imagining two (proper) clocks $\chi_0$ and $\chi_1$, each broadcasting at any time its clock reading together with the chain of signals which has been received from the other clock right at the emission time. Of course, this activity of the transmitters does not rely on any information about possible users receiving the signals.
In the figure we are representing only the signals received by the user $(t_c, x_c)$ which, as we said above is an event, not a worldline.
If the user were a worldline it would receive all signals broadcasted by clocks at different times. However, the position of the user as well as the position, orbital parameters and timing of the clocks at the emission of the last signals have to be obtained from these data alone without
relying on data which may be obtained in the past by the user operator. That is what we meant by saying that the rPS is {\it instantaneous}, i.e.~the user is regarded as an event and has to compute everything with data received at once.
}

Before sketching the solution, we need to introduce some notation which will be useful later in higher dimensions when drawing diagrams as in Figure 1 and 2 will become difficult.
First, we shall use the affine structure on Minkowski space, so that the difference $P-Q$ of two points $P, Q\in M$ denotes a vector (tangent to $M$ at the point $Q$ and) leading from 
$Q$ to $P$.
On the tangent space the Minkowski metric induces inner products so that we can define a pseudonorm $(P-Q)\cdot (P-Q)=|P-Q|^2$ so that the vector $P-Q$ is {\it lightlike} iff $|P-Q|^2=0$.

Secondly, if we have $k$ clocks, namely $\chi_0, \chi_1, \dots, \chi_{k-1}$, we shall have an infinite sequence of events along them, namely $p_0, p_1, p_2, \dots$.
Our naming convention will be that the point $p_n$ is along the clock $\chi_i$ iff $n\hbox{\rm\ mod} k=i$ and $p_i$ will be at a later time than $p_{i+k}$.

\begin{figure}[htbp] 
   \centering
  \includegraphics[width=9cm]{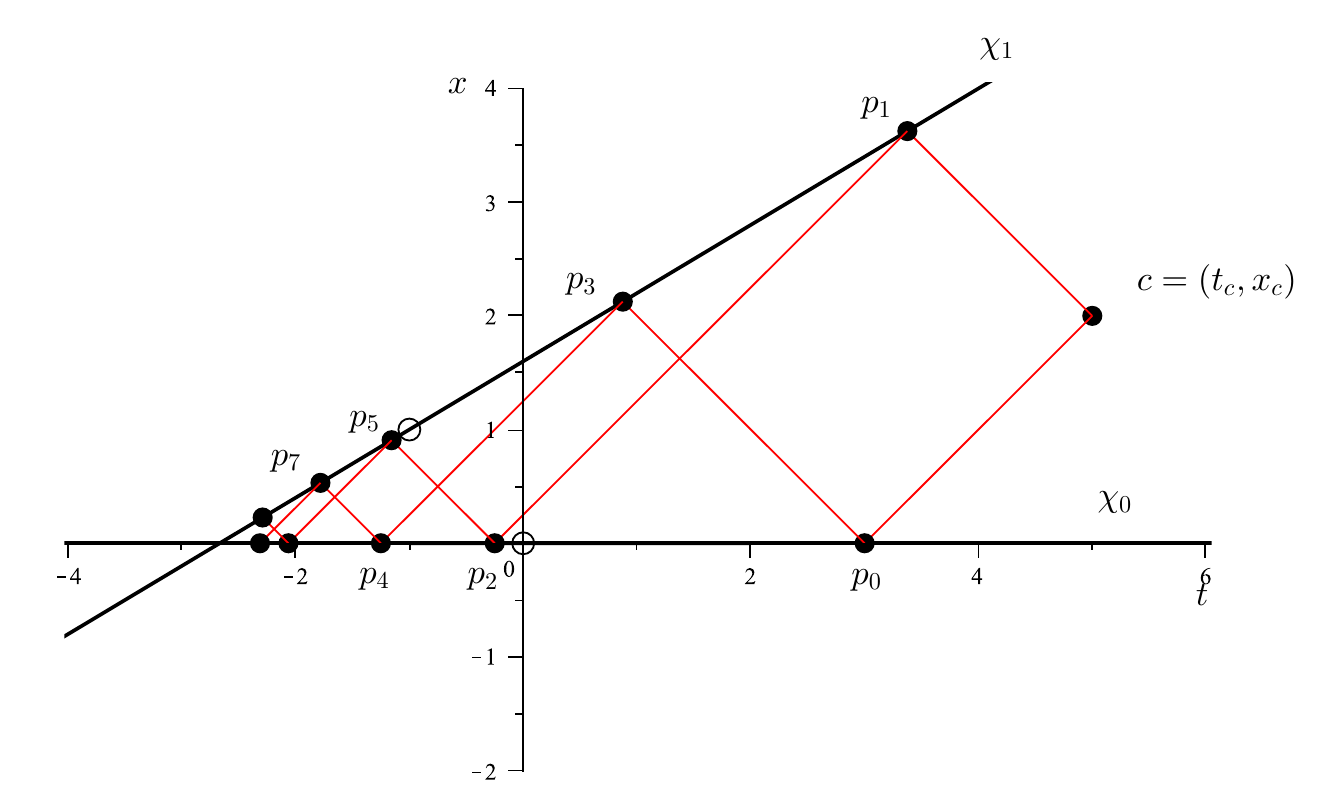} 
   \caption{\small\it Past messages $(s_0, s_1, s_2, s_3, s_4, \dots)$ exchanged from the clocks to the {user in the case of two satellites in 1+1 Minkowski spacetime}.}
   \label{fig:F2}
\end{figure}

Here in Figure $2$ we have $k=2$ so that $p_0, p_2, p_4, \dots$ are events on $\chi_0$, while $p_1, p_3, p_5, \dots $ are events along $\chi_1$.

Then the segments $c-p_0$, $c-p_1$, $p_2-p_1$, $p_3-p_0, \dots$ are all light rays so that one has
\begin{equation}
\begin{aligned}
&|c-p_0|^2=0
\quad
|c-p_1|^2=0
\quad
|p_2-p_1|^2=0
\quad
|p_3-p_0|^2=0
\quad
\dots
\end{aligned}
\label{Eqs}
\end{equation}

The clock reading at the event $p_i$ will be denoted by $s_i$. 
Each clock will be mirroring all signals it receives at $p_i$ from the other clock(s) in addition to the value $s_i$ of its reading at that event. 
Accordingly, the {user} will receive the whole sequence $(s_0, s_1, s_2, s_3, s_4, \dots)$.

\begin{figure}[htbp] 
   \centering
  \includegraphics[width=7cm]{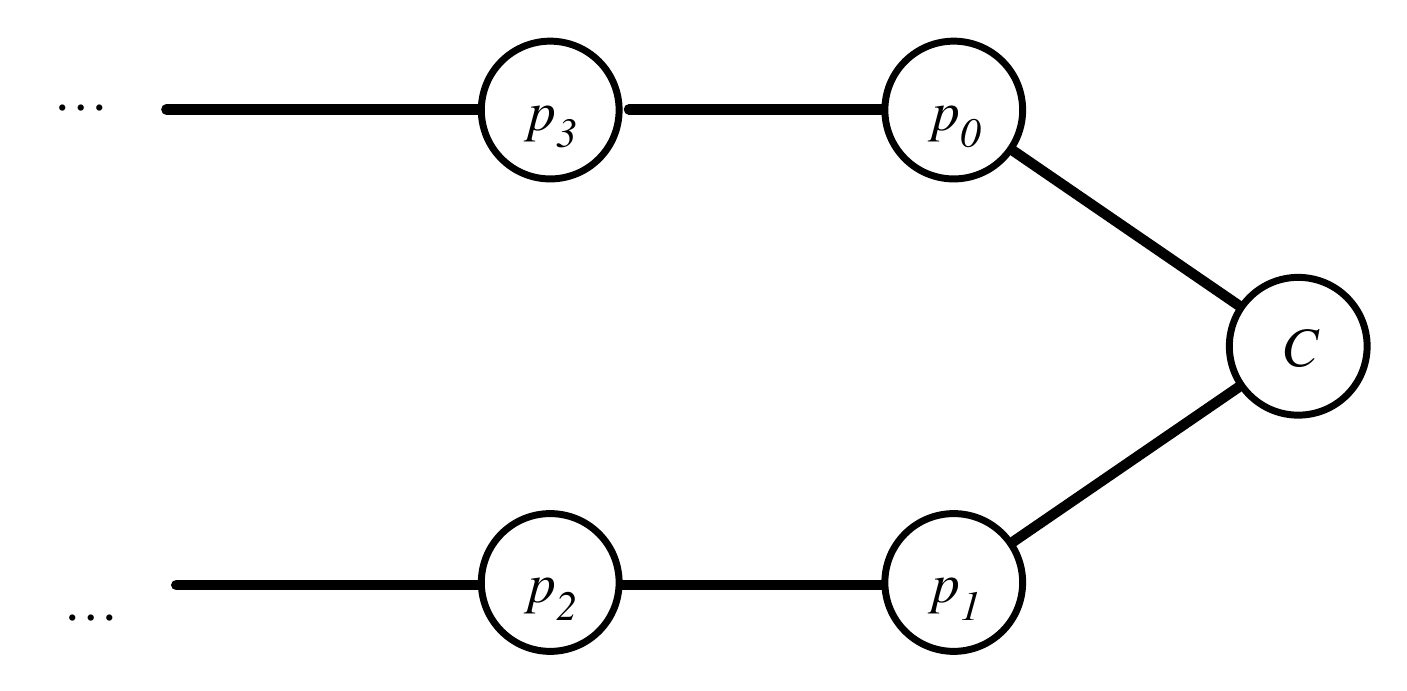} 
   \caption{\small\it Graph representing the {signal exchanges} shown in Figure 2.}
   \label{fig:F3}
\end{figure}

Finally, in 2d, Figure 2 is enough to describe the whole convention setting. However, in higher dimensions such pictures will be difficult to read. For this reason, we replace 
the description by  a graph as in Figure 3.
In these graphs, all lines represent a set of equations such as (\ref{Eqs}).

Now one can check that equations 
\begin{equation}
\begin{aligned}
&|p_2-p_1|^2=0
\quad
|p_3-p_0|^2=0
\quad 
|p_4-p_3|^2=0
\quad
|p_5-p_2|^2=0
\end{aligned}
\label{Eq1}
\end{equation}
admit solutions
\begin{equation}
\al=\al(s_i)
\quad
\ze=\ze(s_i)
\quad
t_1=t_1(s_i)
\quad
x_1=x_1(s_i)
\end{equation}
with $i=0, 1, 2, 3,4,5$ and that one has $\al^2-\ze^2=1$.

Then one can use the first two equations
\begin{equation}
|c-p_0|^2=0
\quad
|c-p_1|^2=0
\label{Eq2}
\end{equation} 
to solve for 
\begin{equation}
t_c=t_c(s_i)
\quad
x_c=x_c(s_i)
\end{equation}

Actually, by solving the system one gets eight solutions. Four of them are spurious solutions since they do not satisfy the equations
\begin{equation}
|p_4-p_7|^2=0
\qquad
|p_5-p_6|^2=0
\label{Eq3}
\end{equation}
The remaining four solutions then identically satisfy all the following equations.

Finally, if our assumptions about free fall are accurate, the remaining equations are identically satisfied.

To be precise one obtains multiple solutions  (as we discussed above we are left with four solutions) but the correct solution can be selected by checking that all the vectors
\begin{equation}
c-p_0,
\quad
c-p_1,
\quad
p_1-p_2,
\quad
p_0-p_3,
\quad
p_3-p_4,
\quad
p_2-p_5,
\quad\dots
\end{equation}
are future directed.
This reduces the solutions to two.

To find the unique solution, we can utilize Poincar\'e invariance. 
We have already used the boost invariance to adjust the clock $\chi_0$. However, we have a residual invariance with respect to spatial reflections.
We could have originally used this to keep $x_1\ge 0$. This condition can now be used to select a unique solution between the two residual solutions of the system.  

In other words, by using the signals $(s_0, \dots, s_5)$ one is able to uniquely determine both the clock parameters $(\al, \ze, t_1, x_1)$ and the {user} position in spacetime $(t_c, x_c)$.
There is no need of clock calibration or synchronisation. 
Of course, this means that the infinite sequence $(s_0, s_1, s_2, \dots)$ is not independent.
Indeed, we can compute allowed sequences in a {\it simulation phase} in which we assume $(\al, \ze, t_1, x_1, t_c, x_c)$ as parameters, and we compute the signals 
$(s_0, s_1, s_2, \dots)$ by using equations (\ref{Eq1}), (\ref{Eq2}), (\ref{Eq3}), \dots

Then we start the {\it positioning phase}, in which we consider those signals as parameters, and we determine the unknowns $(\al, \ze, t_1, x_1, t_c, x_c)$,
so that the positioning phase essentially deals with the inversion of what is done in simulation mode.

{In a sense, the {\it simulation-positioning} paradigm provides a general framework to design and analyse rPS. In the {\it simulation phase} we assume a physical situation, a given gravitational field a given setting of transmitters and {user} and describe the signals exchanged among them.
As a result we are able to determine emission and observation events on  transmitters' worldlines, 
i.e.~the signals which are eventually received by the {users}. As we said, in simulation mode one knows the physical parameters (e.g.~the orbital parameters of the transmitters, additional non-gravitational forces acting on transmitters, maybe the mass generating the gravitational field and so on) and computes the potentially infinite sequence $(s_0, s_1, s_2, \dots)$ of signals received by the {user} at a given event.

In the {\it positioning phase} we go the other way around, we try and invert the correspondence, recovering the information about the physical configuration, only out of the signals received at an event. That is precisely what a rPS {user} should do to determine the configuration of the satellites and forces acting on them first and then its own position in spacetime.

Later on, when discussing the case of a 2d Schwarzschild spacetime, we shall also argue that the  positioning phase can be reduced quite in general to a generic simulation phase. If one is able to perform the simulation phase for generic parameters, then a least square analysis can be used  for positioning. 
}

Let us finally remark that when we prove that the particular system of inertial coordinates $(t, x)$ is allowed, then consequently, any other inertial coordinate system is allowed as well. 

We  can also add an unexpected acceleration $a$ to the clock $\chi_1= (t_1+ \al s, x_1+ \ze s + \Frac[1/2] a s^2)$ while computing the signals to be transmitted to the {user}.
If the {user} does not know about the acceleration and it keeps assuming (wrongly this time) that the clock is free falling, then
one can show that the {user} can still determine the parameters of the clock, but this time the constraint $\al^2-\ze^2=1$ and the redundant equations cannot be identically satisfied.
This shows that the {user} is potentially able to test the assumptions we made and to self-diagnose their break down, {i.e.~it is robust}.

If the acceleration dies out, as soon as the transmitters exchange a few signals the system manages to satisfy the constraints and it becomes operational again.
That shows the rPS to be self-calibrating.

\section{Minkowski in  1+2 and 1+3 dimensions}

When we consider Minkowski spacetime in dimension three the situation becomes more complicated and one needs to think about what is going on in order to apply the simple program we presented in dimension two.

In dimension three we consider three proper clocks $\chi_i$, with $i=0,1,2$. Each has five degrees of freedom, an initial position $(t_i, x_i, y_i)$ and an initial direction
given by $(\alpha_i, \ze_i, \xi_i)$ obeying the constraint $\al_i^2-\(\ze_i^2+\xi_i^2\)=1$.

One can still use the Poincar\'e invariance to set the first clock to be $\chi_0: s\mapsto (s, 0, 0)$, 
though one still has two clocks $\chi_1, \chi_2$ and will have to deal with signals back and forth between them (which will turn out to be coupled quadratic equations, compared with lower dimension cases where each equation contained only the parameters of one clock at a time).
Moreover,  in dimension two one has only two light rays through any event and each of them goes to a clock, while in dimension three one has infinitely many light rays 
through an event and one has to select the one intersecting a clock.

Finally, in 1+1 dimensions when Poincar\'e invariance is used to fix the 0th clock we are left with a discrete residual invariance with respect to spatial reflections.
On the other hand, in 1+3 dimensions, one is left with a 1-parameter rotation group (as well as spatial reflections), i.e.~with $O(2)$, which can be used to set $y_1=0$ and $x_1\ge 0$.

However, one can still show that the unknowns which in our case are now $(t_1, x_1, y_1, \al_1, \ze_1, \xi_1, t_2, x_2, y_2, \al_2, \ze_2, \xi_2)$
can be computed from the signals and the constraints $\al_i^2-\(\ze_i^2+\xi_i^2\)=1$.
The redundancy and the Poincar\'e fixing described above can be used to select a unique solution; the further redundancy is used to check assumptions.

\begin{figure*}[htbp] 
   \centering
  \includegraphics[width=12cm]{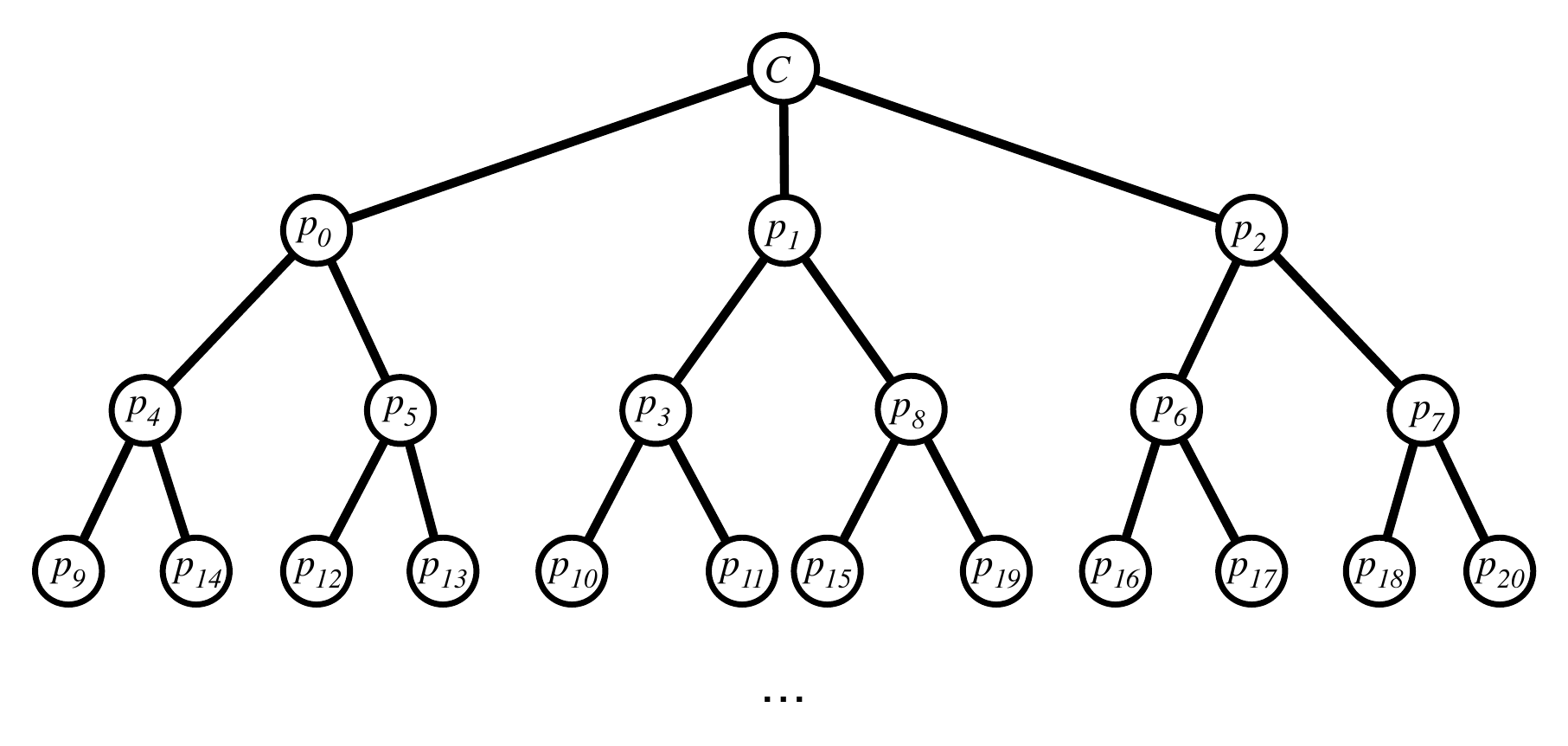} 
   \caption{\small\it Graph representing the signal exchange in 1+2 Minkowski spacetime.}
   \label{fig:F4}
\end{figure*}

To do that, we  used the scheme of signals shown in Figure 4.
We first use the equations
\begin{equation}
\begin{aligned}
|p_0-p_4|^2=0
\quad
|p_1-p_3|^2=0
\quad&
|p_4-p_9|^2=0
\quad
|p_3-p_{10}|^2=0
\quad
|p_6-p_{16}|^2=0
\quad
|p_7-p_{18}|^2=0\\
&
\al_1^2-\(\ze_1^2+\xi_1^2\)=1
\end{aligned}
\end{equation}
which just depend on $(t_1, x_1, y_1, \al_1, \ze_1, \xi_1)$, to determine the parameters of the first clock.
Since three of them are not independent, one has two extra parameters, a sign $\ep_1=\pm 1$ and an angle $\om_1$ which are left undetermined and will be fixed later on.
Similarly, we used equations
\begin{equation}
\begin{aligned}
|p_0-p_5|^2=0
\quad
|p_2-p_6|^2=0
\quad&
|p_5-p_{12}|^2=0
\quad
|p_3-p_{11}|^2=0
\quad
|p_8-p_{15}|^2=0
\quad
|p_6-p_{17}|^2=0\\
&\al_2^2-\(\ze_2^2+\xi_2^2\)=1
\end{aligned}
\end{equation}
which depend on $(t_2, x_2, y_2, \al_2, \ze_2, \xi_2)$, to determine the parameters of the second clock, leaving two parameters, again a sign $\ep_2=\pm 1$ and an angle $\om_2$ undetermined
due to relations of the equations.

Then one has the extra equations representing signals between the clocks $\chi_1$ and $\chi_2$
\begin{equation}
\begin{aligned}
&|p_1-p_8|^2=0
\quad
|p_2-p_7|^2=0
\quad
|p_4-p_{14}|^2=0
\quad
|p_5-p_{13}|^2=0
\quad
|p_8-p_{19}|^2=0
\quad
|p_7-p_{20}|^2=0
\end{aligned}
\end{equation}
Using these equations (and depending on the four possibilities for $(\ep_1, \ep_2)$) we can determine $\om_1-\om_2$,
thus leaving only $\om_2$ undetermined. This fact accounts for the residual Poincar\'e invariance which can be used to set $y_1=0$ and $x_1\ge 0$.
We note that two of the four possibilities for $(\ep_1, \ep_2)$ need to be abandoned because they are incompatible with these equations.

Finally, two solutions are found, only one of which agrees with the gauge fixing $x_1\ge 0$.
Thus, also in this case, the {user} is able to determine the parameters of the clocks uniquely.

Once the clocks are known, one can use the equations
\begin{equation}
|c-p_0|^2=0
\qquad
|c-p_1|^2=0
\qquad
|c-p_2|^2=0
\end{equation}
to determine the {user} position $(t_c, x_c, y_c)$.
In this case, one needs to solve equations on a one by one basis, {in a wisely chosen order}, to control the details of the procedure.

As we see, Minkowski spacetime $M_3$ in 1+2 dimensions is quite different with respect to Minkowski
spacetime $M_2$ in 1+1 dimensions. Already in that simple case one needs to select equations wisely to solve them. We argue that for higher dimensions things do not escalate and higher dimensions are qualitatively as $M_3$.  We shall just sketch these cases since eventually the real cases of physical interest will be Schwarzschild (or Kerr) in 1+3 dimensions. After all we are mainly thinking about a rPS around a star or a black hole, not navigation in interstellar space as in \cite{T1, T2}.

Let us start by considering Minkowski spacetime $M_4$ in 1+3 dimensions.
Let us consider $4$ freely falling (i.e.~inertial), otherwise identical, clocks $(\chi_0, \chi_1, \chi_2, \chi_3)$.
Spacetime has a Poincar\'e invariance which can be fixed by setting $\chi_0$ at the origin 
(i.e.~$\chi_0=(s, 0, 0, 0)$),
$\chi_1$ moving in a spatial plane $x^3=0$, parallelly to the axis $x^1$
(i.e.~$\chi_1(s)=(t_1+\al_1 s, x_1+ \ze_1 s,  c_1, 0)$) with the constraint $\al_1^2-\ze_1^2=1$ and $x_1>0$.
The other two clocks are unconstrained as $\chi_i(s)=(t_i+\al_i s, x_1+ \ze^1_i s, y_1+ \ze^2_i s,
z_1+ \ze^3_i s)$ with the constraint $\al_i^2-(\ze^1_i)^2-(\ze^2_i)^2-(\ze^3_i)^2=1$ and $i=2,3$.

In dimension $m=4$ the Poincar\'e group is of dimension 10, hence the whole system has $18= 7\times 4-10$ degrees of freedom, namely 
it is described by 21 parameters 
\begin{equation}
\begin{aligned}
&t_1, x_1, c_1, \al_1, \ze_1;
\qquad
t_2, x_2, y_2, z_2, \al_2, \ze^1_2, \ze^2_2, \ze^3_2;
\qquad
t_3, x_3, y_3, z_3, \al_3, \ze^1_3, \ze^2_3, \ze^3_3
\end{aligned}
\end{equation}
with 3 constraints.

As far as the signals are concerned, we have 4 signals in emission set $0$, namely the signals from transmitters to the {user}, $4\times 3=12$ in emission set $1$,
$12 \times 3= 36$  in emission set $2$,
$36 \times 3$  in emission set $3$, and so on.

Among them, if we fix two clocks $\chi_i$ and $\chi_j$ with $i\not=j$,
$2$ are exchanged between $\chi_i$ and $\chi_j$, at the $1^{st}$ emission set, 
$6$ at the $2^{nd}$ emission set, $18$ at the $3^{rd}$ emission set, and so on.
Also 4 are sent from a clock to the {user}, at emission set $0$.

Accordingly, we have $2+6=8$ signals exchanged between $\chi_0$ and $\chi_i$ in emission sets $1$ and $2$, for each $i=1,2,3$. 
The associated conditions just depend on the parameters of the clock $\chi_i$ and partially fix them,
as we showed it happens in dimension 3 (i.e.~$2+1$).
Then we have $3\times 8=24$ extra more equations from signals exchanged between clocks $i\not=0$ which can be used later to fix the redundancy.

{The algorithm given above is completely general and can be extended to any number of dimensions. In Appendix A, we collected some considerations about how to extend the model to a Minkowski spacetime in this case.} 

\section{Schwarzschild in 1+1 dimensions}

This Section is an attempt to utilize the procedure described in the previous Sections on a curved spacetime. 
We are not endowing the model with any physical meaning; gravity is sometimes considered trivial in dimension two, since Einstein equations are identically satisfied.
However, probably one could argue for a meaning as radial solutions in  4-dimensional Schwarzschild spacetime.
In fact, the Minkowski cases we studied above are vulnerable to two different concerns:

\begin{itemize}
\item[1)] we extensively used  the affine structure of $\R^n$ to write the equations to identify light rays;

\item[2)] the metric is flat; thus, the Lagrangian for geodesics has an extra  first integral (the conjugate momentum to $x$ which is cyclic).
\end{itemize} 

In both cases, we should check that we are able to perform the computation on a more general curved spacetime, 
otherwise what we have done above would be restricted to SR.

Let us {consider coordinates $(t, r)$ and} try the metric 
\begin{equation}
g= -A(r) dt^2 + \Frac[dr^2/ A(r)]
\qquad\qquad
A(r):= 1-\Frac[2m/r]
\end{equation}
which corresponds to the Lagrangian for the geodesics
\begin{equation}
L=\sqrt{
			A{\( {\Frac[dt/ds]} \)^2}- {{\Frac[1/A]} \( {\Frac[dr/ds]} \)^2}
	 } ds=\sqrt{\Frac[A^2-\dot r^2/A]} dt
\label{Lag}
\end{equation}

There are some reasons to prefer this Lagrangian to the ordinary quadratic one.
First of all, it is invariant with respect to re-parameterisations. The quadratic Lagrangian is not and is valid only when one parametrises with proper time. 
{On the contrary, here in Lagrangian (\ref{Lag}) the parameter $s$ is {\it a priori} arbitrary. It can be specified to the proper time or to the relative time $t$, as well as any other parameter.}
However, the physical motions are represented by trajectories in spacetime, not by parametrised curves.
The parameter along the curve (any parameter, including the proper time) is introduced as a gauge fixing of this invariance
and just to use the variational machinery introduced in mechanics.
Accordingly, a Lagrangian which accounts for re-parameterisations is better than one which does not, exactly as a gauge invariant dynamics is better than 
one which is written in a fixed gauge.

Secondly, we shall use it for both particles and light rays. 
If proper time is an available gauge fixing for particles, it is not for light rays.
Thus, this choice of Lagrangian allows us to discuss light rays on an equal footing with particles.

Thirdly,  one can always fix the gauge later on: the use of this Lagrangian is not a restriction.

The solutions of this Lagrangian are geodesic trajectories.
Of course, one can try and solve its Euler-Lagrange equation analytically, 
although obtaining this solution strongly relies on the specific form of the metric.

Instead, we try and develop a method to find geodesics relying on first integrals and the Hamilton-Jacobi (HJ) method.
This method works on any spacetime which allows separation of variables for the corresponding HJ equation (which are classified; see \cite{HJ,  HJ1}).
Once a complete integral of the HJ equation is known 
(which may be obtained by the method of separation of variables)
then solutions are found ({\it{}only}) by inverting functions.

\subsection{Hamiltonian formalism}

The momentum associated to the Lagrangian (\ref{Lag}) is
\begin{equation}
p= \Frac[\del L/\del \dot r] =  -\Frac[\dot r/\sqrt{A(A^2-\dot r^2)}]
\iff
\dot r= -\Frac[pA\sqrt{A}/\sqrt{1+Ap^2}]
\label{LegTr}
\end{equation}
which corresponds to the Hamiltonian
\begin{equation}
H= - \sqrt{A}\sqrt{1+Ap^2}
\end{equation}

The corresponding HJ equation is
\begin{equation}
 - \sqrt{A}\sqrt{1+A\(W'\)^2}=E
 \iff
 W'= \mp\Frac[\sqrt{E^2-A}/ A]
\end{equation}
where prime denotes the derivative with respect to $r$.
The complete integral for HJ is hence
\begin{equation}
S(t, r; E)= -E t  \mp \int \Frac[\sqrt{E^2-A} / A]dr
\end{equation}

\subsection{The evolution generator}

For later convenience, we would like to express it as a function of the initial condition, i.e.
\begin{equation}
\begin{aligned}
F(t, r; t_0, r_0)=& S(t, r) - S(t_0, r_0)= 
-E \cdot(t-t_0)  \mp \int_{r_0}^r \Frac[\sqrt{E^2-A} / A]dr
\end{aligned}
\label{IntegralSchwarzschild}
\end{equation}
which will be called the {\it evolution generator}, once we eliminate $E$
(see Appendix {B}).

\
The evolution generator contains the information for finding the general solutions of Hamilton equations, i.e.~general geodesic trajectories.
In fact, one has
\begin{equation}
-\Frac[\del F/ \del E] = -\Frac[\del S/ \del E] (t, r)+\Frac[\del S/ \del E] (t_0, r_0)= P-P_0=0
\label{myeq}
\end{equation}
which is zero since the momentum $P$  conjugate to $E$ is conserved,  $S$ being a solution of the HJ equation.

In principle, one could use this equation to obtain $E(t, r; t_0, r_0)$
and replace it above to obtain the evolution generator $F(t, r; t_0, r_0)$.

Once the evolution generator has been determined, we can determine the geodesic trajectory
passing through $(t, r)$ and $(t_0, r_0)$ by computing
{
\begin{equation}
\begin{cases}
p=\Frac[\del F/ \del r]\\
p_0=-\Frac[\del F/ \del r_0]\\
\end{cases}
\end{equation}
}where $p_0$ is the initial momentum to be selected so that the geodesics will eventually pass through $(t, r)$
while $p$ is the momentum when it arrives at $(t, r)$.
Equivalently, one can use the inverse Legendre transform (\ref{LegTr}) to obtain the initial and final velocities.

Accordingly, the flow of the transformations $\Phi_{t-t_0}:(t_0, r_0)\mapsto (t, r)$ is canonical and describes completely 
the geodesic flow.

One can use this method to obtain again the geodesics in Minkowski space (setting $A=1$), this time {in coordinates $(t, r)$ and} without resorting to the affine structure but using the manifold structure only.

In the Schwarzschild case, one can solve the integral (\ref{IntegralSchwarzschild}). However, the resulting equations (\ref{myeq})
turn out to be too complicated to be solved for $E$. Consequently, we need to learn how to go around this issue.
For our Schwarzschild--like solution, i.e.~for $A=1-\Frac[2m/r]$, we can introduce a dimensionless variable $r= 2m\rho$ to obtain
\begin{equation}\label{F_rho}
\begin{aligned}
F(t, r; &t_0, r_0)=   -E\cdot (t-t_0)  
\mp 2m \int_{\rho_0=\Frac[r_0/2m]}^{\rho=\Frac[r/2m]} \Frac[\sqrt{\rho}  \sqrt{{(E^2-1)\rho +1}} / \rho-1]d\rho
\end{aligned}
\end{equation}

The limit to lightlike geodesics is obtained by letting $\dot r\arr \pm {A(r)}$, which corresponds to $p\arr \mp \infty$,
which in turn corresponds to the limit $E\arr -\infty$.

Thus, for light rays, we are interested in the solutions of (\ref{myeq}) which diverge to $-\infty$.
Given a clock $\ga: s\mapsto (t(s),r(s))$ and an event $(t_c, r_c)$,
if we want to determine a light ray going from the clock to the event, we should determine $s=s_\ast$ on the clock so that
there is a lightlike geodesic from $(t, r)=(t(s_\ast),r(s_\ast))$  to $(t_0, r_0)=(t_c, r_c)$. Since the geodesic one determines is lightlike, the corresponding $E(t, r; t_0, r_0)$ diverges.

Even though the explicit form of $E(t, r; t_0, r_0)$ is hard to find we can make the substitution $E=1/\ep$ in the equation (\ref{myeq})
and then take the limit $\ep\arr 0^-$, i.e.~take the limit through negative values of $\epsilon$. 

In the Schwarzschild case, we obtain for (\ref{myeq})
\begin{equation}
(t-t_0) \mp \(r- r_0+2m\ln\(\Frac[r-2m/r_0-2m]\)\)+O(\ep)=0
\end{equation}
the two signs corresponding to ingoing and outgoing geodesic trajectories.
This allows a divergent solution (i.e.~$\ep=0$) iff
\begin{equation}
t-t_0 =\pm\( r- r_0+2m\ln\(\Frac[r-2m/r_0-2m]\)\)
\label{lightrays}
\end{equation}
Once we fix the initial condition $(t_0, r_0)$, this provides an {explicit} definition $t(r)$ of the lightlike geodesics trajectories through it, parametrised by $r$.
Thus, in view of separation of variables, HJ method provides us with an exact, analytical, description of light rays {parameterized in terms of $r$}.

Moreover, before taking the limit to $E\arr -\infty$, Equation \eqref{F_rho} is also a good description of {test} particles which we can use to describe the motion of transmitters.
For $-1<E\le 0$ one has bounded motions, while for $E\le -1$ one has unbounded motions.

The bounded motions have a maximal distance they reach before falling in again.
This is obtained by conservation of $E$ as the value of $r=r_M$ such that
\begin{equation}
E^2= A(r_M)
\end{equation}
and then one has directly the two branches of the motion as 
\begin{equation}
\dot r^2= A^2\(1-\Frac[A/E^2]\)
\iff
t-t_M= \mp E \int^r_{r_M} \Frac[dr/{A}\sqrt{E^2-A}]
\end{equation}
This suggests to use $r$ as a parameter along each branch. Notice that since we have not used any expansion or approximation the result we have obtained is analytic and exact. {Similarly}, any unbound motions (either ingoing or outgoing) can be fully described by the parameter $r$.

{Let us point out that the parameterisation of light rays in terms of $r$ is physically odd. One should prefer a parameterisation in terms of relative time $t$ instead, since, of course, proper time is not defined on light rays. However, such a parameterisation $r(t)$ relies on the inversion of equation (\ref{lightrays}) which of course formally exists although it is hard to obtain explicitly in practice. Thus here is where we take advantage of reparameterisation invariance and use a suitable parameter.}

Suppose now {we} fix two transmitters (e.g.~a bound clock $\chi_0$ and an unbound outgoing clock $\chi_1$).
{ We already know in Minkowski the user needs to be in between the transmitters, otherwise the coordinates become degenerate since a second satellite on one side does not add information for determining the position of the user. Similarly, also here in Schwarzschild, we set a user in between the transmitters.} 

We can trace back light rays exchanged by the clocks and eventually to the {user}, obtaining what is shown in Fig.~5.
{This is obtained as in Minkowski; we consider a moving point along the worldline of the transmitter and move it until we find a solution of the equation (\ref{lightrays}) of light rays. This determines an event $p_n$ on the clock worldline and a value of $r_n$ of the parameter for it.
As we discussed the parameter $r_n$ is not a physical time, hence we need to transform it to the corresponding proper time $s_n$ along the clock. Again, this would not be necessary if we used a clock parameterisation in terms of its proper time in the first place. However, such a parameterisation relies on the form of the inverse function of $\tau(r)$,
which is not explicitly known. Using $r$ as a parameter allows us to avoid formal inversions of functions and keep the model explicitly computable.}

In this way, we are able to find {\it exactly} the points $(p_0, p_1, p_2, p_3, \dots)$ at which the signals are emitted by the clocks and the corresponding clock readings
$(s_0, s_1, s_2, s_3, \dots)$.
In other words, we can, also in this case, exactly model the simulation phase in the 2d Schwarzschild case.

If the {user} does not assume the correct gravitational field and is instructed to find its position 
anyway, the constraints turn out to be violated. 
In principle, the {user} can say that the transmitters are not moving as they would be expected in Minkowski space.
Thus system is {robust}.

However, in Schwarzschild spacetime, the positioning phase is much more difficult to be performed.
We have a system of {(trascendent)} equations to be solved and, of course, we can check that the parameters used in simulation modes do verify them.
However, we cannot show without actually solving the system, for example, that the solution is unique or that we are able to select it among the solutions as the real {user's} position.

Let us take the opportunity to explain a different strategy to solve the system which shows quite generally that one does not actually need to  solve the system in positioning system,
provided we can perform the simulation phase efficiently, for generic enough parameters.

The idea is to transform the solution of a system 
{
\begin{equation}
\begin{cases}
f_0(\al_0, \al_1, \dots, \al_n)=a_0\\
f_1(\al_0, \al_1, \dots, \al_n)=a_1\\
 \dots \cr
f_k(\al_0, \al_1, \dots, \al_n)=a_k\\
\end{cases}
\end{equation} 
}where $(a_0, a_1, \dots, a_k)$ are the signals predicted in simulation mode {for the unknown model parameters $\al_i$},
into a search for the minima of an auxiliary function $\chi^2$ 
{
\begin{equation}
\chi^2 (\al_0, \al_1, \dots, \al_n) =  \sum_{i=0}^k \(f_i(\al_0, \al_1, \dots, \al_n)-a_i\)^2
\end{equation}
}chosen so that  the minima of $\chi^2$ are achieved exactly on the solutions of the system.

There are quite a number of tools developed to find minima since that is used for fits.


{We used MultiNest (see \cite{MN1,MN2, MN3}), 
a powerful Bayesian inference tool developed for {a} highly efficient computation of the evidence by producing and analysing the posterior samples from the distributions with 
(an unknown number of) multiple modes and pronounced degeneracies between the parameters.
Relying on the posterior distribution provided by the software we are able to detect the presence of more than one solutions and calculate them.
Hence to solve the system we just need to be able to compute the functions
{$f_i(\al_0, \al_1, \dots, \al_n)$}  for arbitrary values of the parameters
{$(\al_0, \al_1, \dots, \al_n)$}, which is what we learnt to do in positioning phase and then MultiNest is able to explore the parameter
space to look for minimal and best fit values, which for us
are best approximations of the solutions of the system.
After that, as we said previously, one can check that there are many modes (as it happens
when more solutions are present) by just analysing the posterior distribution.}
 
{\bf In the Schwarzschild case, for two clocks with parameters $\chi_0= (t_0=40, {r}_0=4, v_0= -\Frac[3/8])$ and $\chi_1= (t_1=-40,  {r}_1=4, v_1= \Frac[17/48])$ where $(t_0,r_0)$ and $(t_1,r_1)$ are the events at which the clocks are re-set and $v_0$ and $v_1$ are their radial velocities at the  re-set. 

The first transmitter is falling in ($v_0<0$), the second going out ($v_0>0$). Setting the central mass $m=1$ and a {user} at $c=(t_c=60,  {r}_c=15)$ we can compute the first 16 signals:}
\begin{equation}
\begin{aligned}
& s_0=1.159156591
\quad
s_1=71.51766019 \\
&
s_2=-11.26675255
\quad
s_3=46.96574042\\
& s_4=-21.45717682
 \quad s_5=41.97205382\\
&
 s_6=-23.59547256
 \quad
  s_7=37.62526214\\
 &
 s_8=-25.46320683
\quad s_9=36.70632744\\
 &
 s_{10}= -25.85844196
\quad s_{11}= 35.90306640\\
 &
 s_{12}= -26.20396026 
\quad s_{13}= 35.73306272\\
 &
 s_{14}= -26.27709250
\quad s_{15}= 35.58442173\\
\end{aligned}
\end{equation}

The corresponding posterior distributions generated by MultiNest are shown in the triangular plots in Figure 6.
{The first triangular plot used only the first 10 signals, the second used 16 signals.
In both cases MultiNest found one mode (solution) only, determining the unknown parameters correctly.}

We can see that the solution is unique and that we have a {localisation which agrees with the parameters used in simulation. Notice that here we are working with a minimal model setting (no transmitter redundancy, no perturbation, no clock errors, no clock drifting with respect to proper time) in a quite extreme gravitational field (definitely in a strong regime).
Although the localisation in this simple model is not very accurate, it serves as a proof-of-concept. 
We should also remark that even discussing accuracy here would not be very meaningful since we see no way to compare accuracy in this model 
(in 1+1 dimensions, in the strong regime, no real orbits around the central mass) with more realistic settings (in dimension 4, in the weak regime, with transmitters orbiting the central mass). 
Of course, further investigation need to be devoted to accuracy estimate of localisation as well as the domain of the coordinates.
}

{Also}, we have not optimised anything here, the {user} can restrict heuristically the region to scan for solutions using its past positions and one can tune precisions to improve localisations, or drop MultiNest for a simpler minimiser if one is not interested in posterior distributions.

{Let us finally remark that, in} this example, we can be nasty and not inform the {user} about  the actual value of $m$, leaving it free to be fitted, 
that the clocks  were unbounded, that the clock $\chi_0$ was ingoing and the clock $\chi_1$ was outgoing.
{One can show that } this information is still obtained by the fit result.

\onecolumngrid

\begin{figure}[htbp] 
   \centering
   \includegraphics[width=12cm]{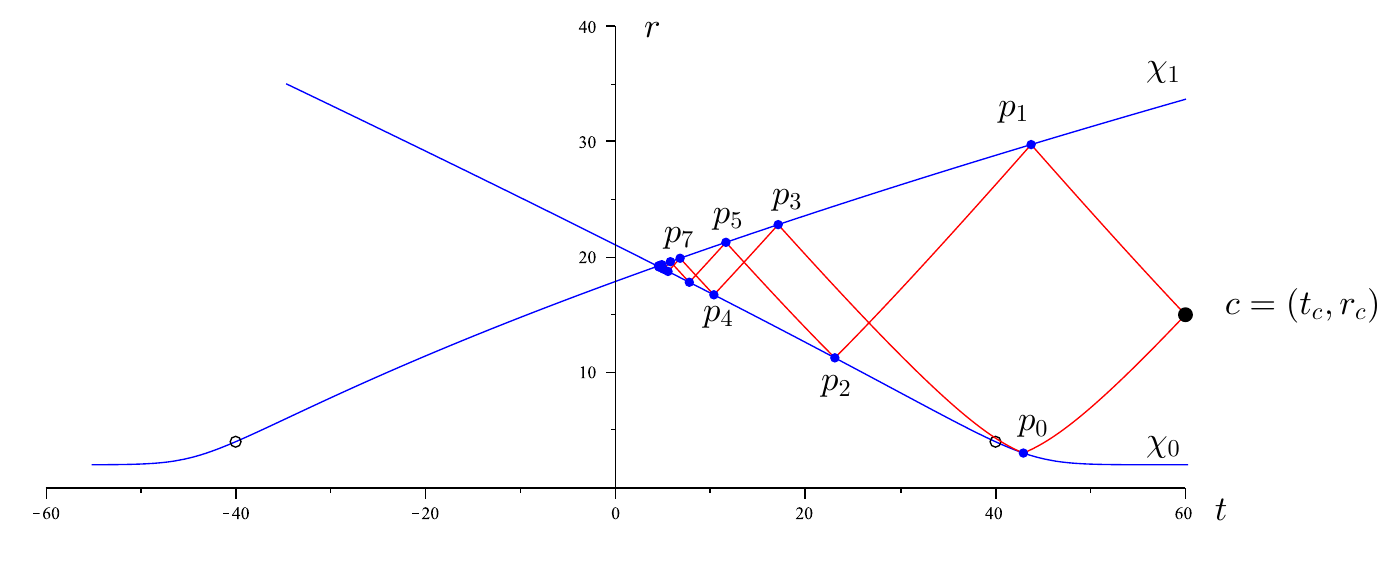} 
   \caption{\small\it Simulation phase for 2d Schwarzschild. {Empty dots are reset events.} }
   \label{fig:Schwar}
\end{figure}

\section{Conclusions and perspectives}

We showed that one can define a rPS system without resorting to rulers and synchronisation at a distance
so that it is simple, instantaneous, discrete, self-calibrating, and {robust} in the sense defined above.
We considered cases in 1+1, 1+2 and 1+3 dimensions, flat or curved spacetimes.

This setting perfectly integrates with the EPS framework and axioms as well as with the framework introduced by Coll and collaborators.
As it happens in EPS, everything is produced by starting from the worldlines of particles and light rays.
In view of what we proposed above, we can also define coordinates (in addition to conformal structure and projective structure) which 
are a better bridge with the conventions used, e.g., in physics.

If Coll and collaborators focus on positioning, in this paper we considered the problem from a different perspective.
From a foundational viewpoint, it is generally recognised that GR is the most fundamental layer of our description of classical phenomena. 
Since in most experiments we need to use coordinates and  positioning of events,
 we should be able to do that {\it before} we start investigating the detailed properties of spacetime, of physical fields, and the evolution of the universe.

From this viewpoint rPSs have an important foundational relevance, since they are a prerequisite to experiments.
The more they are considered fundamental the less detail can be used to design them.
In particular, being self-locating and self-calibrating are important characteristics just because in principle they do not require that we model
the motion of satellites from mission control.
This is obtained by considering the parameters governing  clocks (their initial conditions {and frequencies}) as unknown parameters
instead of fixing them as control parameters.

As the unknown parameters grow in number, one clearly needs more data to solve for them. 
The available data can be increased by different strategies. 
We can add clocks (since the unknown parameters grow linearly with the number of clocks, while the exchanged signals grow quadratically) or we can go back in time considering and mirroring signals exchanged by the clocks.

However, adding the first signals exchanged by the clocks is  insufficient to solve for all unknown parameters.
Coll et collaborators directly resorted to a continuous flow of data which also simplifies the analysis.
This approach relies on the inversion of functions which is notoriously problematic in general. 
We instead showed that one can go back in discrete steps as described in Figure 2 and 5,
keeping the sequence of signals discrete and using a finite sequence to 
solve for the unknown parameters and the others as constraints to check accuracy of the assumptions.

{Our design of rPS is based on discretisation of signals, which is not a new idea in the literature. For example, refs.~\cite{Cadez} and \cite{RovelliGPSb} as well as reference 6 of ref.\cite{Tarantola09} use some form of discretisation as well.
Although our design has similarities with them, especially related to discretisation, it is also true that there are differences. 
In general, we do not assume that the orbital parameters are known {\it a priori}, e.g.~by solving for the geodesic equations in a given metric. We instead use information on the differential timing of the satellites along their worldline.
We determine these parameters in a process which involves all the clocks of the satellite constellation whereas the method in \cite{Cadez} requires a knowledge of the satellite orbits. In addition,  in the approach of \cite{Cadez} users are considered as worldlines (hence they could use information accumulated in the past) whereas in our case they are simple events and currently they use only information available at the moment in which they receive broadcasted signals from satellites. Of course, in future works, we shall possibly use past information to optimize calculation, e.g.~by restricting the parameter space of the  of the satellite motion. This will most likely increase efficiency.
For a user in the more complicated case of a Schwarzschild spacetime, our procedure leads to the exact determination for $t(r)$ exploiting not only Hamiltonian methods but also a complete use of the parameterization invariance, which as far as we know has not been fully employed yet. }

One advantage in our design is that the positioning itself uses just the first few emission sets of signals.

For example, in a rPS around the Earth with astrometric parameters 
similar to NAVSTAR-GPS, we saw that look back time corresponds to the $2^{nd}$ emission set
of signals. NAVSTAR-GPS satellites are in orbit at about $25 Mm$, hence two satellites are at most $6\cdot 10^{7}\> m$.  
In this design, a signal that can be used for positioning can bounce twice on satellites and then be redirected to the {user}, for a total of $1.5\cdot 10^{8}\> m$, which at the speed of light is way less than a second. Thus we expect that a rPS on Earth could have a look back time of about a second.
{Of course this is a rough estimate, which in order to be made precise, would need that first a realistic model of rPS in dimension 4 in conditions similar to Earth is produced.
Accordingly, we are not claiming here anything precise. However, we are  remarking that in some experimental conditions, 
{although not in general (see \cite{Tarantola09}),}
there can be a split of data into a part which is relevant for positioning and a part which is relevant for gravitometric. This splitting, if it exists, relies on assumptions, e.g., about the gravitational field in terms of symmetries and the associated conserved quantities, which is clearly impossible in general, although we expect them may be valid in restricted conditions.
For example, trivially, if we are very far away from the central mass then, by asymptotic flatness, we know that Minkowski is a good approximation, and it seems reasonable to expect that one can do positioning in that approximation, even though observing data long enough eventually one should be able to see curvature effects, anyway. 
Of course, further investigation is needed in this direction to show if our expectation are met and can be made precise.}
 
{In a self-calibrating rPS,  signals which are older than the look back time} are used only for checking the assumptions, i.e.~for measuring whether the gravitational field agrees with what is assumed. Accordingly, one can also argue that perturbations of the gravitational field become relevant for positioning only when they are measurable within $1s$.
We believe this setting {may turn out to be} a good compromise between simplicity of design and effectiveness.

{Of course, there is a lot of work to be done in the direction of having a realistic rPS. As we said one needs to account for the fact real atomic clocks do not beat proper time but some (known) function of it. This does not seem to be an issue since we parameterised everything in terms of $r$ including proper time.
Hence all that seems to be needed is applying such a time drift function to the signals.
However, this will affect the propagation of the errors and the final accuracy.

Similarly, of course the Earth's gravitational field is not spherically symmetric. One sooner or later will have to add multipolar corrections; see \cite{Gomboc}. Also for this reason, perturbations need to be introduced, for positioning or gravitometric purposes. We did not consider this issue here so a further study needs to be devoted to them.

Also Earth's atmosphere has not been discussed at all while its effects have to be accounted before rPS can be proposed as a realistic positioning system.
}

Again from a foundational viewpoint it is interesting to note, and worth further investigation, that rPSs potentially can be used to measure the gravitational field.
Theoretically, they can be used in a gravitational theory, possibly different from standard GR, as a tool to produce observable quantities. Since they are well integrated 
with the generally covariant framework, they are candidates to pinpoint differences between different theories on an observational grounds,
namely to design experiments to compare gravitational theories. 

{For a practical viewpoint instead, one needs a way of estimate precision of positioning as well as characterising the region where the positioning defines a well-defined coordinate system. The two issues are strongly related since uncertainties in coordinates spoil them in practice.
Of course, that can be done by simulation, although some insights could be obtained by using the theory of characteristics and caustics developed for geometric optics.
Further studies will be devoted to discuss precision in a more realistic situation.
} 


\begin{figure*}[htbp] 
   \centering
 \includegraphics[width=16cm]{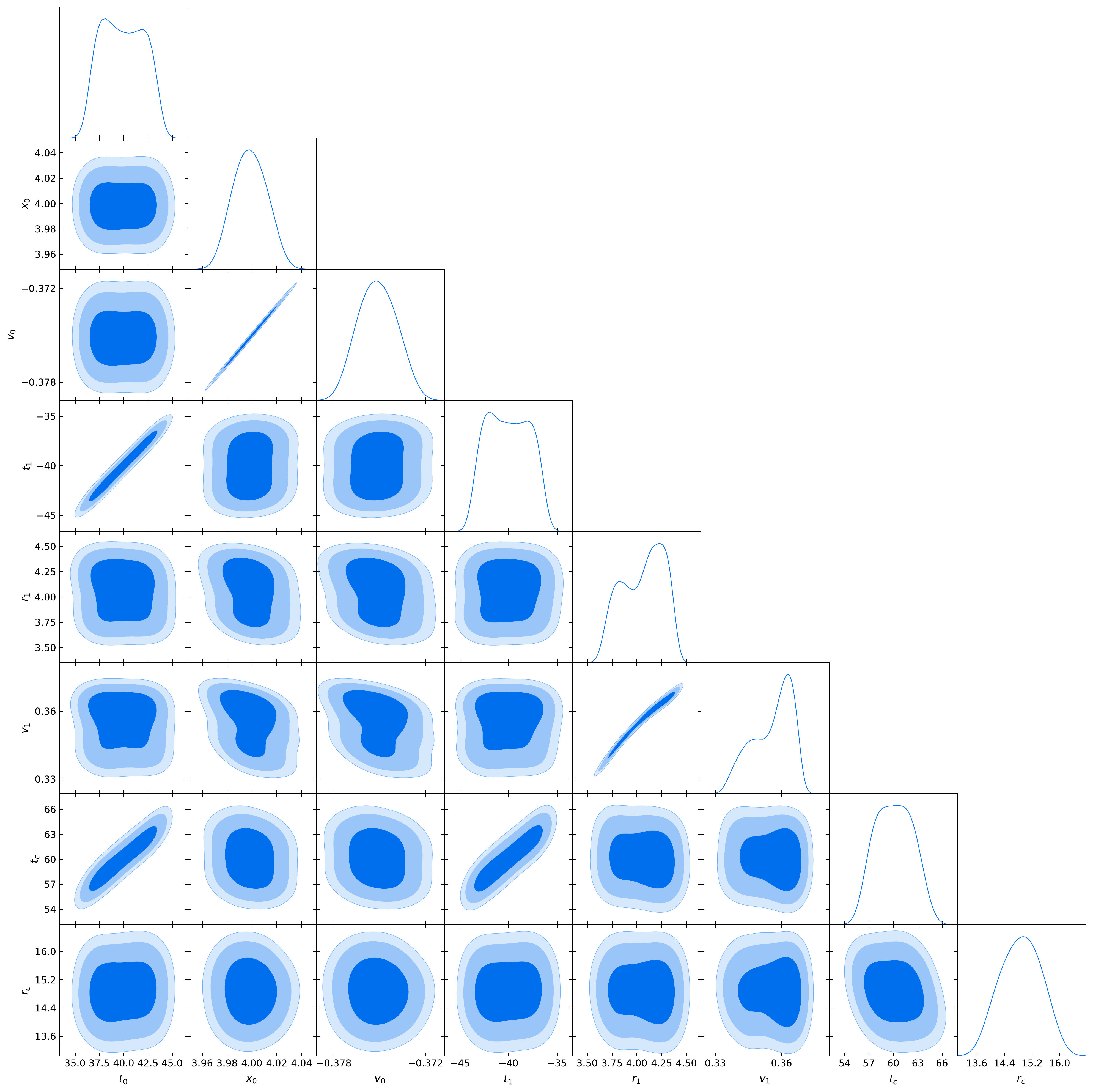} 
   \caption{\small\it Triangular plots for two different simulations starting with the same parameters.
   The mass $m$ is fixed to $m=1$.
   The plot uses ten signals. }
\end{figure*}
\begin{figure*}[htbp] 
   \centering
 \includegraphics[width=16cm]{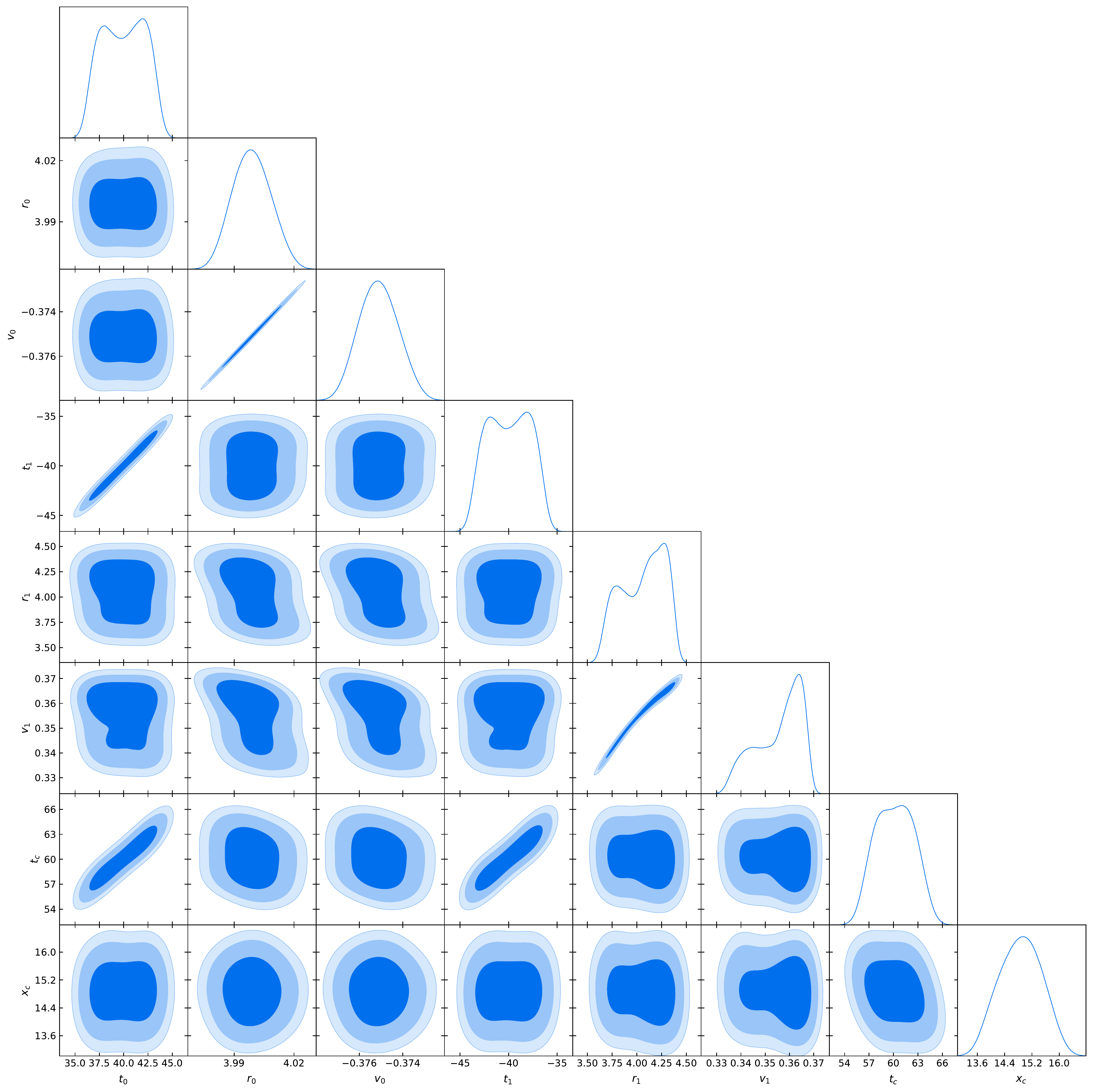} 
   \caption{\small\it Triangular plots for two different simulations starting with the same parameters.
   The mass $m$ is fixed to $m=1$.
   The plot uses sixteen signals.}
   \label{fig:TriangularPlots}
\end{figure*}

{\small \begin{acknowledgments}
This article is based upon work from COST Action (CA15117 CANTATA), supported by COST (European Cooperation in Science and Technology).

We also acknowledge the contribution of INFN (Iniziativa Specifica QGSKY), the local research project {\it  Metodi Geometrici in Fisica Matematica e Applicazioni (2015)} of Dipartimento di Matematica of University of Torino (Italy). This paper is also supported by INdAM-GNFM.

S. Carloni was supported by  the {\it Funda\c{c}\~{a}o para a Ci\^{e}ncia e Tecnologia} through project 
IF/00250/2013 and partly funded through H2020 ERC Consolidator Grant ``Matter and strong-field gravity: New frontiers in Einstein's theory'', grant agreement no. MaGRaTh-64659.

L. Fatibene would like to acknowledge the hospitality and financial support of the Department of Applied Mathematics, University of Waterloo where part of this research was done.
This work was supported in part by a Discovery Grant from the Natural Sciences and Engineering Research Council of Canada (R. G. McLenaghan)
\end{acknowledgments}
}

\vskip10pt

{

\section*{Appendix A: Minkowski in dimension $m$ and lookback time}

In a Minkowski spacetime of dimension $m$ we consider $m$ proper clocks $(\chi_0, \dots, \chi_{m-1})$.
We fix Poincar\'e invariance setting the first clock to be $\chi_0=(s, 0, \dots, 0)$.
We are left with a spatial residual invariance parametrised by $O(m-1)$ which we shall need to fix the gauge.

Each clock receives $(m-1)$ signals from the other clocks and the graph analogous to that in Figure 4 becomes of order $m$.
In fact, in the graph representing the messages exchanged in dimension $m$, each node will receive $m-1$ edges, each representing an incoming message (and an equation to be satisfied) 
and emit one edge representing the message broadcast by that clock.
The node representing the {user} is exceptional, since it receives $m$ signals from the clocks and it does not emit, hence appearing as the root of the graph.

Thus the graph accounts for
\begin{equation}
\si:=m+ m(m-1) + m(m-1)^2= m(m^2-m+1)
\end{equation}
signals. To each of these signals there is an associated equation.

The signals exchanged between $\chi_0$ and $\chi_i$ are in fact $2m$ (for any $i=1..m$) which, together with the constraint $\al_i^2- \( \ze_{i1}^2+ \ze_{i2}^2 +\dots+ \ze_{i (m-1)}^2\) =1$,
partially fixes the parameters of the clock $\chi_i$.

Then one has $2m$ equations for any pair $(\chi_i, \chi_j)$ to freeze the relative parameters. 
Thus one has
\begin{equation}
\begin{aligned}
&m+ 2m \binomial{m}{2}= m+ 2m \Frac[m(m-1)/2]=
m(m^2- m+1)\equiv \si
\end{aligned}
\end{equation}
signals as discussed above.
The {user} position can be computed once the parameters of the clocks have been determined.

Accordingly, we see that the actual positioning is determined by emission set up to 
the $2^{nd}$.  Earlier signals are then predicted and can be used to check that the assumptions (the kind of metric is assumed, the clocks are freely falling, and so on) are accurate.
Whatever, transient effect there may be before it, it does not affect positioning, it just reveals the transient effects.

For that reason, the rPS has a characteristic time, namely the time at which $2^{nd}$ emission set signals begins (or more generally signals which are actually used for positioning), which is called {\it look back time}. 
Signals used from positioning are after look back time, hence whatever happens before is irrelevant for positioning. In particular, whatever hypotheses one can make (e.g.~gravitational field is static, is spherically symmetric, or whatever else) must be verified after the look back time. If the gravitational field is slowly changing and it is approximately constant after the look back time, then a static metric is a good approximation.

The signals from before the look back time are used for checking the hypotheses done on clocks and gravity. If transient effects happens there, the system fail to validate them. In a sense rPS are detectors of the gravitational field from before the look back time.
}

{
\section*{Appendix B: The evolution generator}

\def\del{\partial}

The evolution generator we defined above already appeared in
Synge\cite{WorldFunction}  (who acknowledges an early appearance in Ruse\cite{WorldFunction2, WorldFunction3}, as well as in Synge\cite{WorldFunction4} himself)
where it is called the {\it world function} and denoted by  $\Omega(P, P')$.
This world function is defined as the arc-length of geodesics and it is defined as the integral
$$
\Om(P, P')=\int_P^{P'} \sqrt{|g_{\mu\nu}u^\mu u^\nu|} \> ds
$$
computed along the geodesics joining $P$ and $P'$.

Since its introduction, the world function has been used by different authors (see \cite{WorldFunction5, WorldFunction6}) essentially as a generating function of geometry,
showing that the main objects of Riemannian geometry can be obtained and related to the derivatives of the world function.
However, in \cite{WorldFunction5, WorldFunction6} not much is said about the actual origin of such function, its relation with the physics of test particles, and why it should be expected to encode most of the geometry of spacetime.
Moreover, $\Om(P, P')$ is treated as an implicit object, with the exception of Minkowski spacetime for which it can be easily given in explicit form.

To the best of our knowledge, the symplectic origin of the world function has been acknowledged much later  (see Benenti\cite{Benenti})
when optics in Euclidean spaces has been studied as an application of generating families. These families turn out to be the Euclidean version of the world function.
Without giving too much details, we can say that the evolution generator is essentially a generating function for a flow of canonical transformations, which can be used to represent the evolution of a Hamiltonian system. In other words the evolution of a Hamiltonian system can always be represented {as}  a flow of canonical transformations, which is driven by a globally Hamiltonian vector field related to the generating function.  This point of view can be applied in the simple {case of} the free particle in a space (or spacetime) as well as any other Hamiltonian system.

Although this setting does not add much in practice, it has two advantages:
first, we know that the generating function of evolution is related to complete integrals of the Hamilton-Jacobi equation
for the existence of which we have a solid theory and many results in literature.

For example, one can define a functional as the action functional computed along a solution $\hat \ga(t)=(q(t), \dot q(t))$ of  equation of motion which joins positions $(t_0,q_0)$ and $(t, q)$, namely  $S(t, q, t_0, q_0)= \int_{(t_0, q_0)}^{(t, q)} L|_{\hat \ga}\> dt$.
It can be shown quite directly and generally, that when it is varied with respect to the endpoint, i.e.~along a deformed solution $\hat q(t)\simeq q(t) + \de q -\dot q \de t $, one has
\begin{equation}
\de S = p(\de q - \dot q \de t) + L \de t
= p\de q - H\de t
\end{equation}
which shows that $p= \frac{\del S}{\del q}$ and that $S(t, q, t_0, q_0)$ is a solution of Hamilton-Jacobi equation 
\begin{equation}
\frac{\del S}{\del t} + H =0
\end{equation}
The second advantage is that this approach shows that the geodesic problem is a special case of a much more general problem: given a Hamiltonian systems, find solutions going from one position $P$ to another $P'$, which is a kind of classical {\it propagator}.

Hence on one side  Benenti's generator of evolution is a complete integral of Hamilton-Jacobi equation corresponding to {the} action integral evaluated along solutions; on the other side, when written in terms of the invariant Lagrangian (\ref{Lag}), it corresponds exactly the functional length of geodesics i.e. Synge's world function.

This last property is related to the classical problem in control theory for Hamiltonian systems. The evolution generator, once regarded as a generating function,
depends on the initial and final positions $(q, q')$ and its derivatives determines which initial momentum $p$ is needed to actually arrive in $q'$
and with which momentum $p'$ one arrives there.

Of course, a complete integral of Hamilton-Jacobi equation depends on a number of first integrals (corresponding to Morse family parameters in Benenti) which we eliminate by writing them as a function $E(t, r; t', r')$ of positions.
This is done formally by solving equation (\ref{myeq}), in practice by taking the limit $E\arr -\infty$ to obtain light rays.
Let us remark that using the Lagrangian (\ref{Lag}), which is invariant with respect to any change of parameterisations, allows us to deal at once with test particles (which have a well define proper time) and light rays (for which proper time will be degenerate). Notice that the fact that light rays can be defined as limit of test particles is in fact one of the axioms (the compatibility axiom) in EPS framework\cite{EPS}. In this sense the EPS approach fits naturally with the symplectic and world function frameworks.

}

\section*{Bibliography}

\twocolumngrid



\end{document}